\documentstyle[twocolumn,epsf]{jpsj}

\catcode`\@=11

\def\simle{\mathrel{\mathpalette\@versim<}}   
\def\simge{\mathrel{\mathpalette\@versim>}}   
\def\@versim#1#2{\lower2.5pt\vbox{\baselineskip0pt \lineskip-.5pt
   \ialign{$\m@th#1\hfil##\hfil$\crcr#2\crcr\sim\crcr}}}

\newcommand{\bequ}{ \begin{equation} }
\newcommand{\eequ}{ \end{equation} }
\newcommand{\barr}{ \begin{array} }
\newcommand{\earr}{ \end{array} }
\newcommand{\beqarr}{ \begin{eqnarray} }
\newcommand{\eeqarr}{ \end{eqnarray} }

\newcommand{\baralpha}{ \begin{eqnal} \beqarr}
\newcommand{\earalpha}{ \eeqarr \end{eqnal}}

\catcode`\@=12

\def\runtitle{
}
\def\runauthor{Yukitoshi {\sc Motome} and Nobuo {\sc Furukawa}}

\def\sf{\rm}

\title{
A Monte Carlo Method for Fermion Systems\\
Coupled with Classical Degrees of Freedom
}

\author{Yukitoshi {\sc Motome} and Nobuo {\sc Furukawa}$^{1}$ 
}
\inst{
Department of Physics, Tokyo Institute of Technology,
Oh-okayama 2-12-1, Meguro-ku, Tokyo 152-8551 \\
$^{1}$Department of Physics, Aoyama Gakuin University, 
Chitose-dai 6-16-1, Setagaya, Tokyo 157-8572
}
\recdate{
}

\abst{
A new Monte Carlo method is proposed for fermion systems
interacting with classical degrees of freedom.
To obtain a weight for each Monte Carlo sample
with a fixed configuration of classical variables,
the moment expansion of the density of states by Chebyshev polynomials
is applied instead of the direct diagonalization of the fermion Hamiltonian.
This reduces a cpu time to scale as
$O(N_{\rm dim}^{2} \log N_{\rm dim})$
compared to $O(N_{\rm dim}^{3})$  for the diagonalization
in the conventional technique;
$N_{\rm dim}$ is the dimension of the Hamiltonian.
Another advantage of this method is that
parallel computation with high efficiency is possible.
These significantly save total cpu times of Monte Carlo calculations
because the calculation of a Monte Carlo weight
is the bottleneck part.
The method is applied
to the double-exchange model as an example.
The benchmark results show that it is possible
to make a systematic investigation using a system-size scaling
even in three dimensions within a realistic cpu timescale.
}

\kword{
Monte Carlo method, moment expansion, Chebyshev polynomials,
parallel computation, double-exchange model
}

\begin{document}
\sloppy
\maketitle

\section{Introduction}
\label{Sec:Introduction}

Lattice fermion models interacting with classical degrees of freedom
have a wide range of application.
In many realistic cases, localized spins or lattice distortions
for instances
are handled as classical degrees of freedom
under the adiabatic approximation.
In this work, we are interested in the cases
where these interactions are primarily important and
many-body interactions between fermions can be neglected.
Although models are simplified in these cases,
they are still very useful to discuss physical properties
in some realistic systems.
For instance, many experimental results in perovskite Mn oxides
have been successfully explained by the double-exchange (DE) model 
with classical localized $t_{2g}$ spins.
\cite{FurukawaPREPRINT}

In such systems, the Hamiltonian is given by
${\cal H}( \{ x_{i} \} )$ and
the partition function is written as
\begin{equation}
	\label{Zdef}
	Z = {\rm Tr_{C}} {\rm Tr_{F}} \exp \left( -\beta \left[
	{\cal H} \left( \{ x_{i} \} \right) - \mu \hat{N}
	\right] \right),
\end{equation}
where ${\rm Tr_{C}}$ and ${\rm Tr_{F}}$ are traces over
classical degrees of freedom denoted by classical variables
$\{ x_{i} \}$ and fermion degrees of freedom, respectively.
Here, $\beta$ is the inverse temperature, $\mu$ is the chemical 
potential and $\hat{N}$ is the particle-number operator.

A prescription to calculate the partition function
in eq.\ (\ref{Zdef}) is Monte Carlo (MC) sampling
on configurations of classical degrees of freedom.
In this method, fermion degrees of freedom are traced out 
beforehand as follows.
The Hamiltonian ${\cal H} \left( \{ x_{i} \} \right)$
defined for a fixed configuration of classical variables $\{ x_{i} \}$
can be represented by $N_{\rm dim} \times N_{\rm dim}$ matrix
where $N_{\rm dim}$ is proportional to the system size
because there is no many-body interaction between fermions.
Then the trace over fermion degrees of freedom is easily
calculated by the diagonalization of the Hamiltonian matrix as
\begin{eqnarray}
	& &
	{\rm Tr_{F}} \exp \left( -\beta \left[
	{\cal H} \left( \{ x_{i} \} \right) - \mu \hat{N}
	\right] \right)
	\nonumber \\
	&=&
	\prod_{\nu=1}^{N_{\rm dim}} \left[ 1+ \exp \left( -\beta \left(
	E_{\nu} \left( \{ x_{i} \} \right) - \mu
	\right) \right) \right],
	\label{logS}
\end{eqnarray}
where $E_{\nu}$ are eigenvalues of the Hamiltonian
${\cal H} \left( \{ x_{i} \} \right)$.
Hence the problem is formally considered as a classical one
in which the partition function can be written as
\begin{equation}
	\label{Zclassical}
	Z={\rm Tr_{C}} \exp
	\left[ -S_{\rm eff}\left(\{ x_{i} \}\right) \right]
\end{equation}
with the effective action
\begin{equation}
	\label{Seff}
	S_{\rm eff}\left(\{ x_{i} \}\right) =
	-\sum_{\nu=1}^{N_{\rm dim}} \log \left[ 1+ \exp \left( -\beta \left(
	E_{\nu} \left( \{ x_{i} \} \right) - \mu
	\right) \right) \right].
\end{equation}
Trace over classical variables
in eq.\ (\ref{Zclassical}) is effectively calculated by a MC sampling.
The MC update is performed using the Boltzmann weight
of the configuration $\{ x_{i} \}$ which is given by
$\exp \left[ -S_{\rm eff}\left(\{ x_{i} \}\right) \right]$.
Configurations of classical variables $\{ x_{i} \}$ are updated
by using the Metropolis algorithm.
Because eq.\ (\ref{logS}) is positive definite, we do not have
a negative sign problem in this fermionic MC calculation.

Actual numerical calculation along the above conventional procedures
is time-consuming and serious to handle large-sized systems.
The bottleneck is the numerical diagonalization
of the Hamiltonian ${\cal H} \left( \{ x_{i} \} \right)$.
It costs a cpu time scaling as $O(N_{\rm dim}^{3})$.
If there are classical degrees of freedom proportional to
$N_{\rm dim}$, as actually seen in many systems,
one MC sweep to update all the values of $\{x_{i} \}$
totally scales as $O(N_{\rm dim}^{4})$.
Let us consider the DE models for instance.
We are especially interested in three-dimensional cases
as realistic models for Mn oxides.
According to our estimation described later, a cpu time
for $1,000$ MC sweeps in $6\times6\times6$-sites system
costs about 10 days and that
in $8\times8\times8$ costs about 10 months
on a standard workstation.
This makes it extremely difficult to make
a systematic size-scaling analysis,
especially in three dimensions.
So far, MC studies have been performed
mainly in one and two dimensions
without the system-size scaling in many cases.
\cite{Yunoki1998a,Dagotto1998}

In this work, we propose an alternative technique of
MC calculations which is of great advantage to reduce the cpu time.
In our method, the numerical diagonalization of the Hamiltonian
is replaced by a moment expansion of the density of states
by using orthogonal polynomials.
This moment expansion costs a cpu time scaling as
only $O(N_{\rm dim}^{2} \log N_{\rm dim})$ 
compared to $O(N_{\rm dim}^{3})$ in the diagonalization.
Moreover,
the procedure can be done by parallel computations
whereas the parallelization is difficult for the diagonalization
in the conventional method.
Efficiency of the parallelization is very high
especially for large-sized systems.
These significantly save a total cpu time for MC calculations
by accelerating the bottleneck part to obtain the MC weights.

The paper is organized as follows:
In \S \ref{Sec:Algorithm}, the new algorithm is introduced.
It is applied to DE models in \S \ref{Sec:Application}
to demonstrate its advantages.
The benchmark results for a cpu time are also shown
in \S \ref{Sec:Application}.
Sec. \ref{Sec:Summary} is devoted to summary.

\section{Algorithm}
\label{Sec:Algorithm}

In the conventional MC technique
for the class of models considered here,
as mentioned in \S \ref{Sec:Introduction},
the Hamiltonian is numerically
diagonalized for each configuration of $\{ x_{i} \}$
to give all the eigenvalues.
The effective action $S_{\rm eff}$ is exactly calculated
through eq.\ (\ref{Seff}) by using these eigenvalues.
However, the exact eigenvalues contain more than enough information
for the purpose of performing practical MC calculations:
It is sufficient to know the density of states
within required accuracy.

In our algorithm, instead of the eigenvalues of the Hamiltonian,
the density of states is estimated 
by using a moment expansion with orthogonal polynomials.
This technique has been originally formulated
to handle huge-sized matrix numerically.
\cite{Wang1994,Silver1994}
The Chebyshev polynomials are convenient in the moment expansion,
which are defined recursively by
\begin{equation}
\label{Trecur}
T_{m+1}(x)=2xT_{m}(x)-T_{m-1}(x),
\end{equation}
with $T_{0}=1$ and $T_{1}=x$ for $-1\le x\le1$.

First, we define a renormalized Hamiltonian $X(\{x_{i} \})$
whose eigenvalues are $-1\le \varepsilon_{\nu}\le 1$
by ${\cal H} = a X + b$ with
\begin{equation}
\label{a&b}
a=(E_{\rm max}-E_{\rm min})/2, \ \ 
b=(E_{\rm max}+E_{\rm min})/2,
\end{equation}
where $E_{\rm max}$ ($E_{\rm min}$) is a highest (lowest)
eigenvalue of the Hamiltonian ${\cal H}$.
Then, the $m$-th Chebyshev moment of the density of states
$D(\varepsilon)\equiv \frac{1}{N_{\rm dim}}
\sum_{\nu}\delta(\varepsilon-\varepsilon_{\nu})$
is defined by
\begin{equation}
\label{mudef}
\mu_{m} = \int_{-1}^{1} T_{m}(\varepsilon) D(\varepsilon) d\varepsilon.
\end{equation}
Once the moments $\mu_{m}$ are calculated, the density of states is
inversely obtained by
\begin{equation}
\label{DOS}
D(\varepsilon)=\frac{1}{\pi\sqrt{1-\varepsilon^{2}}}
\biggl[ \mu_{0} + 2\sum_{m\ge1} \mu_{m}T_{m}(\varepsilon) \biggr].
\end{equation}
The expectation value of an operator $A$ is calculated by
\begin{equation}
\label{<A>}
\langle A \rangle
\equiv
\int_{-1}^{1}A(\varepsilon)D(\varepsilon) d\varepsilon
=
\mu_{0}\nu_{0} +
2\sum_{m\ge1} \mu_{m}\nu_{m},
\end{equation}
where the moments of $A$ is given by
\begin{equation}
\label{momA}
\nu_{m}=\int_{-1}^{1}\frac{d\varepsilon}{\pi\sqrt{1-\varepsilon^{2}}}
A(\varepsilon)T_{m}(\varepsilon).
\end{equation}
In particular, the effective action in eq.\ (\ref{Seff}) is obtained by
\begin{equation}
\label{calcSeff}
S_{\rm eff} = \mu_{0}s_{0} + 2\sum_{m\ge1} \mu_{m}s_{m},
\end{equation}
where
\begin{equation}
\label{smdef}
s_{m}=-\int_{-1}^{1}\frac{N_{\rm dim}d\varepsilon}
{\pi\sqrt{1-\varepsilon^{2}}}
\log\left[1+e^{-\beta\left(a\varepsilon+b-\mu\right)}\right]
T_{m}(\varepsilon)
\end{equation}
with the coefficients $a$ and $b$ in eq.\ (\ref{a&b}).
Eqs.\ (\ref{DOS}) and (\ref{<A>}) are straightforwardly confirmed
by using the orthogonality of Chebyshev polynomials.
\cite{Wang1994,Silver1994}

From the definition (\ref{mudef}), the moments of the density of states
are calculated by
\begin{equation}
\label{calcmu}
\mu_{m}=\frac{1}{N_{\rm dim}}
{\rm Tr} \left\{ T_{m}(X) \right\}
=\frac{1}{N_{\rm dim}} \sum_{\nu=1}^{N_{\rm dim}}
\langle\nu| T_{m}(X) |\nu\rangle,
\end{equation}
where $|\nu\rangle$ are convenient complete basis of the Hamiltonian
 ($\langle\nu_{1}|\nu_{2}\rangle=\delta_{\nu_{1} \nu_{2}}$).
If we obtain the vectors $|\nu; m\rangle \equiv T_{m}(X) |\nu\rangle$,
the calculation of a moment is a vector product;
\begin{equation}
\mu_{m}=\frac{1}{N_{\rm dim}} \sum_{\nu=1}^{N_{\rm dim}}
\langle\nu;0|\nu;m\rangle,
\end{equation}
which costs a cpu time scaling as $O(N_{\rm dim})$.
The vectors $|\nu; m\rangle$ are calculated recursively
by using the relation (\ref{Trecur}) as
\begin{equation}
\label{vecrecur}
|\nu;m+1\rangle=2X|\nu;m\rangle-|\nu;m-1\rangle.
\end{equation}
For a sparse Hamiltonian, the matrix-vector product in eq.\ (\ref{vecrecur})
costs a cpu time scaling as only $O(N_{\rm dim})$.
Furthermore, we use recursive relations of Chebyshev polynomials;
\begin{equation}
T_{2m} = 2T_{m}^{2}-1, \ 
T_{2m+1} = 2T_{m}T_{m+1} - T_{1},
\end{equation}
which enable us to obtain moments up to $M$-th order
from those only up to $M/2$-th order.
A total cpu time to compute the moments
$\mu_{m}$ up to the order of $m=M$
scales as $O(N_{\rm dim}^{2}M)$.

The present method becomes `exact', that is, equivalent to
the direct diagonalization of the Hamiltonian
in the conventional technique
when we take the summations in eqs.\ (\ref{DOS}) and (\ref{<A>})
up to infinite order.
In actual calculations,
we approximate the summations in eqs.\ (\ref{DOS}) and (\ref{<A>})
by finite summations up to $m=M$.
The approximation by the truncation at a finite value of $M$
is controllable
since we can always estimate the errors
through comparison with `exact' results obtained
by the conventional technique
within the range of system sizes we are interested in.
In particular, the truncation error of the effective action,
$\Delta S_{\rm eff}$,
which is a crucial quantity in MC updates,
becomes exponentially small as a function of $M$,
as shown in \S \ref{Sec:Application} for DE models as an example.
This comes from the fact that
the moments $s_{m}$ in eq.\ (\ref{smdef})
become exponentially small to the value of $m$.
This justifies our approximation even for small values of $M$.

We discuss here
$N_{\rm dim}$ dependence of the truncation number $M$
which keeps $\Delta S_{\rm eff}$
(the truncation error of $S_{\rm eff}$) small enough
to perform MC calculations practically.
This is crucial for a cpu time.
From the definition (\ref{Seff}),
$\Delta S_{\rm eff}$ consists of
the sum of errors for each eigenvalue $E_{\nu}$.
Hence, if these exponentially-small errors are statistically
independent of each other,
$\Delta S_{\rm eff}$ is proportional to $\sqrt{N_{\rm dim}}$.
In the case of correlated errors, 
$\Delta S_{\rm eff}$ can be proportional to $N_{\rm dim}$.
For either case, therefore, to estimate $S_{\rm eff}$
within required accuracy,
the necessary truncation number $M$ should be proportional
to $\log N_{\rm dim}$
because $\Delta S_{\rm eff} \sim \exp (-M)$.
This indicates that a total cpu time to obtain moments
for actual MC calculations scales as
$O(N_{\rm dim}^{2} \log N_{\rm dim})$.
These properties of the truncation errors will be examined
for DE models as an example in the next section.

The original formulation of this moment-expansion technique
has been proposed by sampling a few random basis
for the sum in eq.\ (\ref{calcmu}).
\cite{Wang1994,Silver1994}
This reduces a cpu time to scale as $O(N_{\rm dim}MI)$,
where $I$ is the number of the random basis.
However, this sampling method works well
only for huge-sized matrix for which $(N_{\rm dim}I)^{-1/2}$
is small enough.
In our MC calculations, since the matrix sizes for the systems
we are interested in are not so large
($N_{\rm dim} \simeq 10^{3}$), errors are not small and
accumulate through MC updates to lead wrong samplings.
From this, we take a sum over complete basis in eq.\ (\ref{calcmu}).

Another important point to reduce a cpu time
is that the calculations of the moments in eq.\ (\ref{calcmu})
are completely independent for each basis $|\nu\rangle$.
This enables us to compute the sum over $\nu$
in a parallel fashion.
In this algorithm, data to be communicated
between different nodes are small
compared to calculations in each node:
A cpu time to calculate the moments on each processor
is proportional to $N_{\rm dim}^{2}M/N_{\rm PE}$
whereas a time to communicate the calculated moments to add up
is proportional to $M N_{\rm PE}$.
Here, $N_{\rm PE}$ is the number of processors
used in parallel calculations.
This indicates that the parallel computation
is performed very efficiently
as far as $(N_{\rm dim}/N_{\rm PE})^{2}$ is large.
High efficiency of this parallelization
is demonstrated in the last part of the next section.
Since it is difficult to make an efficient parallelization of
the matrix diagonalization when all the eigenvalues are required,
the present algorithm has another advantage in accelerating the
calculation by parallelization.

A cpu time to compute $S_{\rm eff}$ is much shortened
by both the reduction from $O(N_{\rm dim}^{3})$
to $O(N_{\rm dim}^{2} \log N_{\rm dim})$ and
the parallel calculation.
As mentioned in \S \ref{Sec:Introduction},
since the calculation of $S_{\rm eff}$ is the bottleneck
in MC calculations, a total cost is much reduced;
when there are $O(N_{\rm dim})$ classical variables to be updated,
a cpu time for one MC sweep on all these variables scales as
$O(N_{\rm dim}^{3} \log N_{\rm dim}/N_{\rm PE})$ in our algorithm
whereas that in the conventional method scales as
$O(N_{\rm dim}^{4})$.

\section{Application}
\label{Sec:Application}

In this section, we show efficiency of the new algorithm
introduced in \S \ref{Sec:Algorithm}
by an application to DE models
with classical localized spins.
The Hamiltonian is given by
\cite{Zener1951}
\begin{equation}
\label{DEmodel}
{\cal H} = -t\sum_{<ij>, \sigma} \left(c_{i\sigma}^{\dagger} c_{j\sigma}
+ {\rm h.c.} \right) - J \sum_{i} \mbox{\boldmath $\sigma$}_{i}
\cdot \mbox{\boldmath $S$}_{i},
\end{equation}
where $c_{i\sigma}^{\dagger} (c_{i\sigma})$ creates (annihilates)
a $\sigma$-spin electron at site $i$;
$\mbox{\boldmath $\sigma$}_{i}$ is the spin operator
whereas $\mbox{\boldmath $S$}_{i}$
denotes the localized spin at site $i$.
We consider the nearest-neighbor hopping $t$
and the ferromagnetic Hund's-rule coupling $J>0$.
Here the localized spin $\mbox{\boldmath $S$}_{i}$ is
approximated as a classical rotator.
The configuration of the classical rotator is described by
two angles in each site;
$\mbox{\boldmath $S$}_{i} = (\cos\theta_{i} \cos\phi_{i},
\sin\theta_{i} \cos\phi_{i}, \sin\phi_{i})$.
Then, the classical variables $\{x_{i} \}$ in eq.\ (\ref{Zdef}) are
$2N$ variables, $\{\theta_{i}, \phi_{i} \}$ $(i=1,2,\cdot\cdot\cdot,N)$,
in this model (\ref{DEmodel}). $N$ is the number of sites.
The Hamiltonian for a fixed configuration $\{\theta_{i}, \phi_{i} \}$
can be written as $2N\times 2N$ matrix.
Hereafter we take the bandwidth $W=zt\equiv 1$ as an energy unit,
where $z=2D$ is the coordination number
in a $D$-dimensional hypercubic lattice.
Throughout the paper we take $J=4$ which is an
appropriate value to investigate the model in comparison with
Mn oxides experiments.\cite{FurukawaPREPRINT}

This model (\ref{DEmodel}) has been intensively studied
to understand physical properties of perovskite Mn oxides.
\cite{FurukawaPREPRINT}
Some experimental results have been explained; such as
the ferromagnetic metal in carrier-doped region,
the transition to paramagnetic state by increasing temperature and
the negative magnetoresistance near the transition.
As far as we use the conventional MC technique described in 
\S \ref{Sec:Introduction}, it is difficult to study
realistic three-dimensional cases based on the size-scaling analysis,
because of a diverging cpu-time as increasing the system sizes.
So far, numerical studies have been performed mainly in
one and two dimensions.
\cite{Yunoki1998a,Dagotto1998}
In the following, we show that the new algorithm
formulated in \S \ref{Sec:Algorithm} 
reduces a cpu time significantly and makes it possible
to investigate much larger-sized systems
than the conventional technique.

\begin{figure}
\epsfxsize=7.7cm
\centerline{\epsfbox{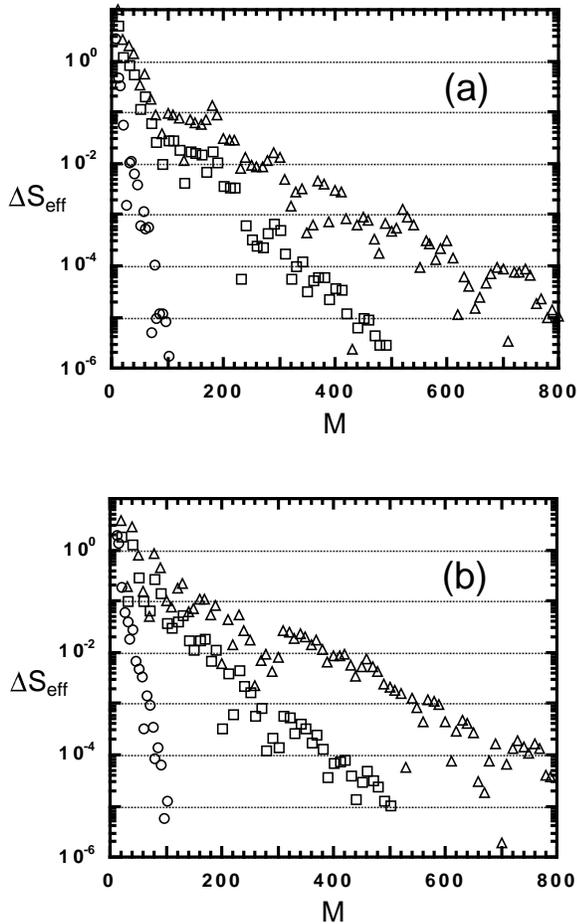}}
\caption{
The error of the effective action
for a configuration of classical variables
when the moment expansion is truncated at $m=M$;
(a) $N=16$ and (b) $N=64$ in one dimension.
We take $J=4$ and $\mu=-4$.
The circles, squares and triangles correspond to the data
for $\beta=10$, $50$ and $100$, respectively.
}
\label{Fig:Seff vs M}
\end{figure}

First, we discuss the truncation error in the moment expansion.
Figure \ref{Fig:Seff vs M} shows the truncation errors of
the effective action
when we truncate the summation in eq.\ (\ref{calcSeff}) at $m=M$.
The errors are estimated as deviations
from the results by the direct diagonalization of the Hamiltonian
in the conventional method.
The errors become exponentially small
to the value of $M$ as shown in the figures.
The exponential decay depends on the temperature.
Approximately, we find a relation;
$\Delta S_{\rm eff} \sim \exp(-M/\beta)$.

\begin{figure}
\epsfxsize=7cm
\centerline{\epsfbox{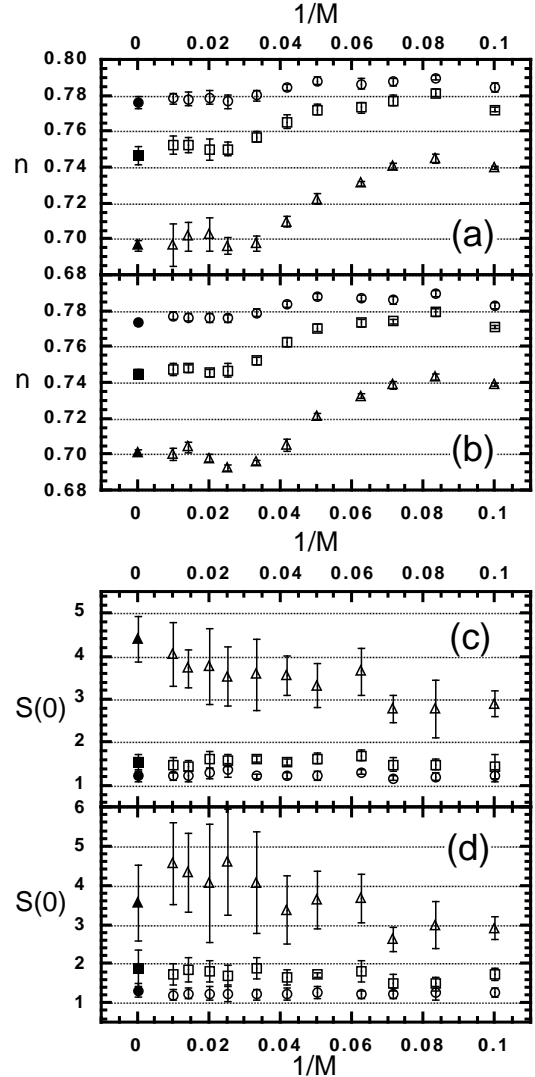}}
\caption{
The truncation $M$ dependence of MC estimations
for physical quantities:
(a) and (b) show the electron density, and 
(c) and (d) show the spin structure factor at the wave number $k=0$.
(a) and (c) are the results for $N=16$, and
(b) and (d) are for $N=64$ in one dimension.
The results are for $J=4$ and $\mu=-3.5$
by $1,000$ MC sweeps on whole lattice sites.
The circles, squares and triangles are for
$\beta=10, 20$ and $50$, respectively.
The filled symbols at $1/M=0$ are the values obtained by
the conventional MC technique
using the diagonalization of the Hamiltonian.
}
\label{Fig:Mdep}
\end{figure}

Although the necessary value of the truncation number $M$
for practical MC is proportional to $\log N_{\rm dim}$
as mentioned in \S \ref{Sec:Algorithm},
actual values of $M$ may depend on models and parameters.
It should be determined in MC results for each case.
We examine here the condition for $M$ by changing temperatures
in the DE model (\ref{DEmodel}).
Figure \ref{Fig:Mdep} shows $M$ dependence of physical quantities.
We calculate here the electron density and the spin structure
factor at the wave number $k=0$ (ferromagnetic component)
by the $1,000$ MC samplings with $S_{\rm eff}$ 
whose moment expansion is truncated at $m=M$.
In all the figures, the values by the conventional technique
with direct diagonalization of the Hamiltonian
are shown at $1/M=0$
because our results should agree with them
in the limit of $M\rightarrow \infty$.
For small values of $M$, the truncation errors are so large that
the estimated values deviate from those by the conventional method.
However, for $M \simge 40$, the expectation values converge
on those by the conventional method within statistical errorbars
in the temperature range of $10 \simle \beta \simle 50$
which we are interested in.
(The ferromagnetic transition temperature is estimated
around $\beta=20$ by the $D=\infty$ technique.
\cite{Furukawa1995})
The behavior is similar between data for different sizes.
This is consistent with the weak $N_{\rm dim}$-dependence
of $M$ ($M\sim \log N_{\rm dim}$) as mentioned
in \S \ref{Sec:Algorithm}.
Therefore, in order to obtain these quantities
of the present model eq.\ (\ref{DEmodel}) within this accuracy,
the value of $M$ can be taken at around $40$
throughout the parameter range of our interests.

\begin{figure}
\epsfxsize=7.7cm
\centerline{\epsfbox{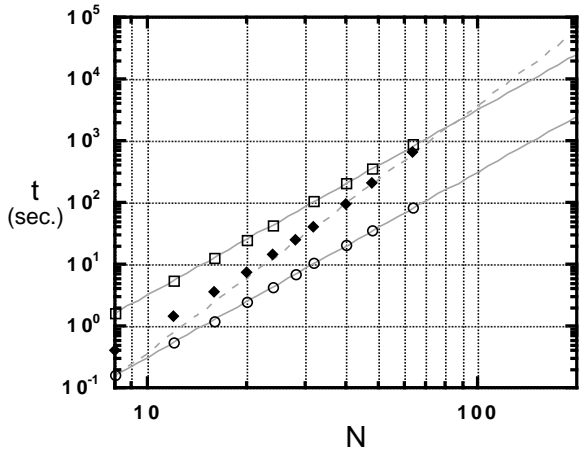}}
\caption{
The benchmark results for both our algorithm and
the conventional one on a workstation
with Alpha-processor 21164 (533MHz).
The results are for $100$ MC sweeps on whole lattice sites.
The filled diamonds are for the conventional technique.
The dotted line is a fit by $N^{4}$.
The circles and squares are for our algorithm with $M=4$ and $40$,
respectively.
The data are fitted by $N^{3}$ as the gray lines.
}
\label{Fig:BM on single}
\end{figure}

Finally, we show the benchmark results for
a cpu time of our algorithm.
Figure \ref{Fig:BM on single} shows the benchmark result
on a workstation with Alpha-processor 21164 533MHz.
A cpu time for MC sweeps on whole lattice sites
in our technique is proportional to $N^{3}$,
whereas the conventional technique costs 
a time proportional to $N^{4}$.
When we take the value of $M$ as $40$
based on the above observation
in Fig. \ref{Fig:Mdep}, our algorithm becomes faster than
the conventional one for $N \simge 80-90$ even in a single CPU.

\begin{figure}
\epsfxsize=7.7cm
\centerline{\epsfbox{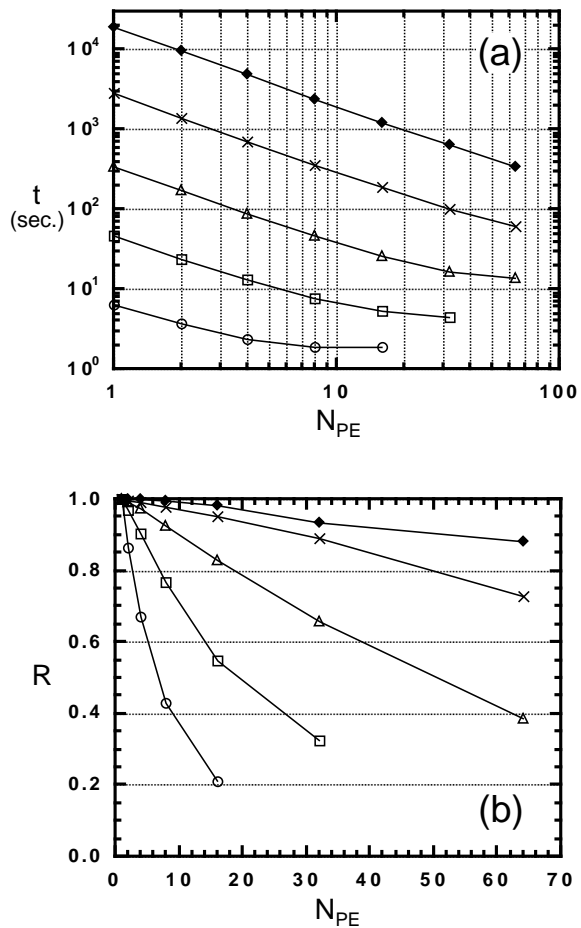}}
\caption{
The benchmark results for our algorithm
on the parallel computer, SR2201 at the computer center of
University of Tokyo.
The results are for $10$ MC sweeps on whole lattice sites
with $M=40$.
The circles, squares, triangles and crosses are
for $N=16, 32, 64$ and $128$ in one dimension, respectively.
The diamonds show the results for
$N=6\times6\times6$ in three dimensions.
(a) is the-number-of-nodes-dependence of a cpu time.
(b) shows the efficiency of the parallelization. See text for details.
The lines are guides to eye.
}
\label{Fig:BM on multi}
\end{figure}

As mentioned in \S \ref{Sec:Algorithm}, moreover,
our algorithm has an advantage to be performed on a parallel computer.
Figure \ref{Fig:BM on multi} shows efficiency of the parallelization.
The parallel calculations have been performed
using SR-2201 at the computer center of University of Tokyo.
We find that in Fig. \ref{Fig:BM on multi} (a),
a cpu time is reduced almost inversely proportional to
the number of processors for large-sized systems.
Figure \ref{Fig:BM on multi} (b) shows the efficiency of
the parallelization defined by
\begin{equation}
R = \frac{1}{N_{\rm PE}}\frac{t(N_{\rm PE})^{-1}}
{t(N_{\rm PE}=1)^{-1}}.
\end{equation}
Here $t(N_{\rm PE})$ is a cpu time on $N_{\rm PE}$ processors.
Note that $R$ takes a value between $0$ and $1$,
and that $R=1$ when the parallelization is perfect.
The parallelization is more effective for larger-sized systems
for the reason mentioned in \S \ref{Sec:Algorithm};
for instance, $R$ remains around $0.9$ for $6\times6\times6$ systems
even when we use $64$ processors.

Table \ref{Table:time} summarizes estimated cpu times
for various system sizes of DE models.
Our algorithm is powerful enough to study three-dimensional systems
with several different sizes within a realistic timescale.
Physical properties in three-dimensional DE models
including the ferromagnetic transition temperature
will be reported elsewhere.

\begin{table}[ht]
\caption{
A cpu time of $1,000$ MC sweeps for various system sizes of DE models
in three dimensions.
The time for the conventional algorithm is estimated
on a work station with a single CPU of
Alpha-processor 21164 (533MHz).
The time for the present algorithm is estimated
on SR2201 at the computer center of University of Tokyo
when we use $64$ processors in parallel.
}
\begin{tabular}{@{\hspace{\tabcolsep}\extracolsep{\fill}}ccc}
\hline
system size & conventional algorithm & present algorithm\\
\hline
$6\times6\times6$ & $\sim 10$ days & $\sim 10$ hours\\
$8\times8\times8$ & $\sim 10$ months & $\sim 6$ days\\
$10\times10\times10$ & $\sim 12$ years & $\sim 6$ weeks\\
\hline
\end{tabular}
\label{Table:time}
\end{table}

\section{Summary}
\label{Sec:Summary}

We have proposed the new algorithm of Monte Carlo calculation
for fermion systems interacting with classical degrees of freedom
under the adiabatic approximation.
The moment expansion of the density of states
by orthogonal polynomials is used,
instead of the direct diagonalization of the Hamiltonian,
to obtain a Monte Carlo weight
for a fixed configuration of classical variables.
Errors of the effective action by the truncation of the expansion
become exponentially small
with increasing the order of the expansion.
This allows us to truncate the expansion at a small order
in practical Monte Carlo calculations.
These reduce a cpu time to scale as
$O(N_{\rm dim}^{2} \log N_{\rm dim})$
from $O(N_{\rm dim}^{3})$ in the diagonalization,
where $N_{\rm dim}$ is the dimension of the Hamiltonian matrix.
This moment-expansion method is controllable
since we can always estimate the truncation errors
through comparison with results by the diagonalization.
As another advantage,
since the moment expansion is independent
for each basis of the Hamiltonian,
our new algorithm can be performed on parallel computers.
Efficiency of the parallelization is very high
especially for larger-sized systems
since data to be communicated between nodes become smaller
compared to calculations in each node.
The reduction of the order of $N_{\rm dim}$ and the parallelization
significantly accelerate 
the procedure to obtain a Monte Carlo weight.
This leads to much reduction of a total cpu time for Monte Carlo
since the calculation of the weight is the bottleneck part.

In order to show the efficiency of our algorithm,
we have applied it to the double-exchange model
with classical localized spins.
We have examined effects of the truncation of the moment expansion
in actual Monte Carlo calculations.
The condition for the truncation is clarified
in the parameter region of interest.
The benchmark results on both a single and multi cpu systems
have been shown.
Our algorithm is a powerful tool to investigate large-sized systems
which are difficult to handle by the conventional technique
within a realistic cpu timescale;
for the double-exchange model, it may make possible
to study three-dimensional systems systematically
through system-size scaling analysis.

There are many other applications of our algorithm;
such as double-exchange models including Jahn-Teller distortion,
\cite{Yunoki1998b}
surface effects in double-exchange systems and
some models for organic compounds.
Our algorithm gives us a possible way to investigate
these problems by using a numerical analysis
for larger-sized systems

\section*{Acknowledgement}

Y. M. acknowledges the financial support of research Fellowships
of Japan Society for the Promotion of Science for Young Scientists.

\end{document}